\documentclass{aa}  

\usepackage{graphicx}

\usepackage{txfonts}

\usepackage{hyperref} 
\hypersetup{
    colorlinks=true,
    linkcolor=blue,
    citecolor=blue,
    filecolor=magenta,      
    urlcolor=blue,
    pdftitle={Overleaf Example}
    }

\begin{document}

   \title{Decoding the cosmological baryonic fluctuations using localized fast radio bursts}

\author{
    Tzu-Yin Hsu \inst{\ref{inst:IOA}, \ref{inst:NTHU}}\fnmsep\thanks{emma30407@gmail.com}
   \and Tetsuya Hashimoto\inst{\ref{inst:NCHU}}\fnmsep\thanks{tetsuya@phys.nchu.edu.tw}
   \and Tsung-Ching Yang\inst{\ref{inst:NCHU}}
   \and Shotaro Yamasaki\inst{\ref{inst:NCHU}}
   \and Tomotsugu Goto\inst{\ref{inst:IOA}}
   \and John Lo \inst{\ref{inst:UCL}}
   \and Po-Ya Wang \inst{\ref{inst:NTHU}}
   \and Yu-Wei Lin \inst{\ref{inst:NTHU}}
   \and Simon C.-C. Ho\inst{\ref{inst:ANU},\ref{inst:CAS},\ref{inst:OzGrav},\ref{inst:ASTRO3D}}
   \and Bjorn Jasper R. Raquel \inst{\ref{inst:PHI}}
}

\institute{Institute of Astronomy, National Tsing Hua University, 101, Section 2. Kuang-Fu Road, Hsinchu, 30013, Taiwan\label{inst:IOA}
  \and Department of Physics, National Tsing Hua University, 101, Section 2. Kuang-Fu Road, Hsinchu, 30013, Taiwan\label{inst:NTHU}
  \and Department of Physics, National Chung Hsing University, 145, Xingda Road, Taichung, 40227, Taiwan \label{inst:NCHU}
  \and Department of Physics and Astronomy, Physics Building, University College London, Gower St, London WC1E 6BT, UK  \label{inst:UCL}
  \and Research School of Astronomy and Astrophysics, The Australian National University, Canberra, ACT 2611, Australia \label{inst:ANU}
  \and Centre for Astrophysics and Supercomputing, Swinburne University of Technology, Hawthorn, VIC 3122, Australia \label{inst:CAS}
  \and OzGrav: The Australian Research Council Centre of Excellence for Gravitational Wave Discovery, Hawthorn, VIC 3122, Australia \label{inst:OzGrav}
  \and ASTRO3D: The Australian Research Council Centre of Excellence for All-sky Astrophysics in 3D, ACT 2611, Australia \label{inst:ASTRO3D}
  \and National Institute of Physics, College of Science, University of the Philippines, Diliman, Quezon City Philippines \label{inst:PHI}
}

\date{Received 23 October 2024; accepted 23 April 2025}

 \abstract
   {}
   {The enigma of the missing baryons poses a prominent and unresolved problem in astronomy. Dispersion measures (DM) serve as a distinctive observable of fast radio bursts (FRBs). They quantify the electron column density along each line of sight and reveal the missing baryons that are described in the Macquart (DM-$z$) relation. The scatter of this relation is anticipated to be caused by the variation in the cosmic structure. This is not yet statistically confirmed, however. We present statistical evidence that the cosmological baryons fluctuate. }
   {We measured the foreground galaxy number densities around 14 and 13 localized FRBs with the WISE-PS1-STRM and WISE $\times$ SCOS photometric redshift galaxy catalog, respectively. The foreground galaxy number densities were determined through a comparison with measured random apertures with a radius of 1 Mpc.}
   {We found a positive correlation between the excess of DM that is contributed by the medium outside galaxies (DM$_{\rm cosmic}$) and the foreground galaxy number density.
The correlation is strong and statistically significant, with median Pearson coefficients of 0.6 and 0.6 and median $p$-values of 0.012 and 0.032 for the galaxy catalogs, respectively, as calculated with Monte Carlo simulations.}
   {Our findings indicate that the baryonic matter density 
outside galaxies exceeds its cosmic average along the line of sight to 
regions with an excess galaxy density, but there are fewer baryons along the line of sight to low-density regions. This is statistical evidence that the ionized baryons fluctuate cosmologically on a characteristic scale of $\lesssim$6 Mpc.}


\titlerunning{Cosmological baryonic fluctuation by localized FRBs}
\authorrunning{Tzu-Yin Hsu et al.}
\maketitle

\section{Introduction}
\label{Intro.}

The most accurate estimation suggests that the combined mass of galaxies, massive groups or clusters, and the warm intergalactic phase amounts to a mere 40$\%$ of the baryons that are predicted by Big Bang theory \citep[e.g.,][]{Shull2012, Fukugita2004, Bristow1994}. The shortfall of the baryonic budget indicates the missing baryon problem \citep{Fukugita1998}.
The intergalactic medium (IGM) stands out as the primary candidate for this quandary.
Nevertheless, the remaining cosmic baryons, that is, missing baryons, exist in environments without significant absorption or emission, and only subtle clues are provided.

Fast radio bursts (FRBs) are enigmatic transients on timescales of milliseconds with mysterious origin \citep[e.g.,][]{lorimer07,petroff22, Bailes2022}. 
One of the observable characteristics of FRBs is their dispersion measure (DM).
The DM represents the column density of ionized electrons along the line of sight, indicating the direct measure of the total number of ionized baryons along lines of sight over cosmologically significant distances. The DM is defined as ${\rm DM}\equiv\int n_{\rm e} ds$, where $n_{\rm e}$ represents the electron number density, and $ds$ represents the line-of-sight element. This property provides a unique means of investigating the distribution of matter in the Universe.

The observed dispersion measure, DM$_{\rm obs}$, is composed of four primary components for a localized extragalactic FRB at redshift $z_{\rm FRB}$,
\begin{equation}
    \textrm{DM}_{\rm obs}=\textrm{DM}_{\rm MW,ISM}+\textrm{DM}_{\rm MW,halo}+\textrm{DM}_{\rm cosmic}+\frac{\textrm{DM}_{\rm host}}{(1+z_{\rm FRB})},
	\label{eq:quadratic}
\end{equation}
with contributions of the Milky Way (MW) interstellar medium (ISM), DM$_{\rm MW, ISM}$, the hot circumgalactic medium in the MW halo, ${\rm DM}_{\rm MW, halo}$, all of the medium outside galaxies, DM$_{\rm cosmic}$, and the host galaxy, DM$_{\rm host}$.
As a tracer of plasma, DM$_{\rm cosmic}$ is proportional to the number of baryons \citep[e.g.,][]{Ioka2003, Inoue2004} in the intergalactic space.

Previous work \citep{Macquart2020} demonstrated the increment of DM$_{\rm cosmic}$ with $z$ (so-called \lq Macquart relation\rq) using localized FRBs. This indicated that the missing baryons reside outside the galaxies. 
The dispersion of the relation is expected to mainly arise from the fluctuation of cosmic baryons, which is still not well understood. If the dispersion is due to the cosmological baryonic fluctuation, the scatter of the Macquart relation might be explained by the projection of the foreground galaxy number density. We note that the Marquart relation has a few outliers, such as FRB20190520B, which shows a large host galaxy contribution to its DM budget \citep{Niu2022}. Therefore, some degree of the fluctuation might be due to the circumburst medium and/or to the host galaxy ISM. This effect could be mitigated by estimating  DM$_{\rm host}$ for individual samples, as outlined in Section \ref{methods}.
Several case studies reported a possible associations between the excess of DM$_{\rm cosmic}$ from the cosmic average ($\langle$DM$_{\rm cosmic}$$\rangle$) and foreground (or in situ) galaxy clusters \citep[e.g.,][]{Conner2023,Lee2023,Wu2023}. 
The DM$_{\rm cosmic}$ along the lines of sight to FRB 20220914A and FRB 20220509G exceeds $\langle$DM$_{\rm cosmic}$$\rangle$, suggesting a DM contribution from the intracluster medium in massive galaxy clusters that host these FRB sources \citep{Conner2023}.
The DM$_{\rm cosmic}$ along the line of sight to FRB 20190520B, which intersects multiple foreground galaxy clusters, also exceeds $\langle$DM$_{\rm cosmic}$$\rangle$ \citep{Lee2023}.
\citet{Wu2023} conducted a stack analysis for 20-30 FRBs whose lines of sight intersect the foreground galaxy halos at $<$ 80 Mpc.
They found a marginal excess in DM$_{\rm cosmic}$ for this sample.
The case studies mentioned above mainly focused on the excess in DM$_{\rm cosmic}$ and on high galaxy-number densities.
However, environments with a low galaxy-number-density should be taken into account to fully understand the cosmological fluctuation of baryons. The foreground galaxy number density around FRBs might be correlated with the deviation of DM$_{\rm cosmic}$ ($\Delta$DM$_{\rm cosmic}$) from the theoretically predicted average value of DM$_{\rm cosmic}$, $\langle$DM$_{\rm cosmic}$$\rangle$ as obtained with the Macquart relation.

We focus on the interesting problem of the scatter in the number of baryons along different lines of sight that is due to density perturbations in the  CMB and Big Bang. This appears as variations in the density of the ionized IGM along different lines of sight and is also expected to appear as variations in the galaxy number density along these lines of sight. 
Therefore, we investigated the hypothetical correlation between $\Delta$DM$_{\rm cosmic}$ and the foreground galaxy number density by systematically studying low- and high-density environments.
We searched for and detected these variations on projected scale lengths of up to $\sim$ 6 Mpc around host galaxies in a small sample of FRBs using data and foreground galaxies with the photo-$z$ catalogs obtained by  WISE $\times$ PS1 \citep{Beck2022} and WISE $\times$ SCOS \citep{Bilicki2016}.
We adopt the Planck18 cosmology \citep{Planck18} as a fiducial model throughout, that is, a cold dark matter cosmology with $(\Omega_{m}, \Omega_{\Lambda}, \Omega_{b}, h) = (0.310, 0.690, 0.04897, 0.677)$, unless otherwise mentioned.

\section{Methods}
\label{methods} 
\subsection{DM excess/deficit calculation}
\label{DM excess}
A DM excess/deficit of an FRB is quantified by a deviation in DM$_{\rm cosmic}$ ($\Delta$DM$_{\rm cosmic}$), which is the subtraction of $\langle$DM$_{\rm cosmic}$$\rangle$ from DM$_{\rm cosmic}$. 
We followed \citet{Macquart2020} to derive $\langle$DM$_{\rm cosmic}$$\rangle$, assuming a {\texttt Planck18} cosmology \citep{Planck18} and a fraction of the cosmic baryons in the diffuse gas of $f_{d} = 0.85$.
DM$_{\rm cosmic}$ was calculated by subtracting the other DM contributions from DM$_{\rm obs}$.
We conducted Monte Carlo simulations to take the uncertainties of DM$_{\rm MW, ISM}$, DM$_{\rm MW, halo}$, and DM$_{\rm host}$ into account.
The observational errors of DM$_{\rm obs}$ are negligibly small compared with the above uncertainties. 
To calculate $\textrm{DM}_{\rm MW,ISM}$, we used a flat probability density function (PDF) centered on the NE2001 model \citep{Cordes2004} estimate with a $\pm20\%$ uncertainty \citep{Cordes2022}.
For $\textrm{DM}_{\rm MW, halo}$, we employed a flat PDF that ranged between 25 pc cm$^{-3}$ to 80 pc cm$^{-3}$ \citep{Prochaska2019, Yamasaki2020}.
DM$_{\rm host}$ was conventionally assumed to be based on either a representative value of the MW disk model \citep[e.g.,][]{Macquart2020} or on cosmological simulations \citep[e.g.,][]{Zhang2020}. 
However, the possible variation in DM$_{\rm host}$ might mimic the measured $\Delta {\rm DM}_{\rm cosmic}$ when a fixed value of DM$_{\rm host}$ is assumed.
To mitigate this issue, we used the observed scattering of individual FRBs to estimate DM$_{\rm host}$ as described below.

Scattering is a time delay in the radio emission through multiple paths within a plasma. This might be a useful tool for measuring DM$_{\rm host}$. 
The scattering time for FRBs in which it can be seen would contain information on DM$_{\rm host}$ because it likely occurs in FRB host galaxies \citep[e.g.,][]{Cordes2022}.
To estimate individual DM$_{\rm host}$, we used the empirical relation between $\textrm{DM}_{\rm host}$ and the scattering time at the 1 GHz ($\tau_{1 {\rm GHz}}$) in the host rest frame \citep{Ramachandran1997,Cordes2022}.
This empirical relation is based on Galactic pulsar observations, but was calibrated for FRBs with the factor$f_{\rm scale}$ \citep{Cordes2022}.
The empirical relation is
\begin{multline}
     \tau_{1{\rm GHz}}({\rm DM}_{\rm host})~({\rm ms})\\
     =1.90\times10^{-7}{\rm DM}_{\rm host}^{1.5}\nu^{-4}(1+3.55\times10^{-5}\,{\rm DM}_{\rm host}^{3.0})f_{\rm scale},
 	\label{eq:tau}
\end{multline}
where ${\rm DM}_{\rm host}$ is in units of pc cm$^{-3}$, and $\nu$ is in GHz. 
We included the scatter of $\tau_{1{\rm GHz}}({\rm DM}_{\rm host})$, $\sigma_{\rm log\tau}$ $=$ 0.76 \citep{Bhat2004,Cordes2019}, in our analysis.
This scatter in turn propagates into the errors of DM$_{\rm host}$ and DM$_{\rm cosmic}$.

We adopted $f_{\rm scale}=3$ to account for the differences in the scattering geometry between cosmological FRBs and Galactic pulsars \citep{Cordes2022}. 
FRB emissions are characterized by plane waves that are due to their long distance to the observer, assuming that the scattering screen is within the host galaxy, while Galactic pulsars are described by spherical wave fronts \citep{Cordes2016}.

Equation \ref{eq:tau} allows us to estimate DM$_{\rm host}$ for a given $\tau_{\rm 1GHz}$.
$\tau_{1{\rm GHz}}$ was derived by $\tau_{\rm 1GHz} = \tau_{\rm obs}(\nu_{\rm obs})^{4}(1+z_{\rm FRB})^{3}$, where $\tau_{\rm obs}$ is the observed scattering at the observed frequency of $\nu_{\rm obs}$.
The $(1+z_{\rm FRB})^{3}$ factor is due to the $\nu^{-4}$ dependence of the scattering time and the $(1+z_{\rm FRB})$ dependence of the observed time.
The signal-to-noise ratio (S/N) of $\tau_{\rm obs}$ is defined as $\frac{\tau_{\rm obs}}{\tau_{\rm err}}$, where ${\tau_{\rm err}}$ is the uncertainty of $\tau_{\rm obs}$. When $\tau_{\rm obs}$ was measured with an S/N $>$ 3, we used its lower and upper errors ($\tau_{\rm err}^{-}$ and $\tau_{\rm err}^{+}$, respectively) and the scatter around Equation \ref{eq:tau} to derive the lower and upper uncertainties of DM$_{\rm host}$ ($\delta {\rm DM}_{\rm host}^{-}$ and $\delta {\rm DM}_{\rm host}^{+}$, respectively).
In our Monte Carlo simulations, the 
lower and upper errors were treated separately
to describe the PDF of DM$_{\rm host}$ with the standard deviation of $\delta {\rm DM}_{\rm host}^{-}$ on one side and $\delta {\rm DM}_{\rm host}^{+}$ on the other side. 
When the S/N of $\tau_{\rm obs}$ was $<$ 3, we used the 3 $\sigma$ upper limit on $\tau_{\rm obs}$ to derive the upper limit on DM$_{\rm host}$ (DM$_{\rm host}^{\rm upper}$). In this case, we adopted a flat PDF of DM$_{\rm host}$ ranging from 0 to DM$_{\rm host}^{\rm upper}$ pc cm$^{-3}$.
A fraction of FRBs only includes the upper limit on $\tau_{\rm obs}$. 
The DM$_{\rm host}$ of these samples were treated in the same manner as $\tau_{\rm obs}$ with an S/N $<$ 3.
At the time of writing, no scattering time measurement for FRB 20200906A, FRB 20211127I, FRB 20211212A, and FRB 20220105A was available. 
In these cases, we assumed a Gaussian PDF of DM$_{\rm host}$ centered at DM$_{\rm host}=100$ pc cm$^{-3}$ with a standard deviation of 50 pc cm$^{-3}$.
The analysis of these samples can be improved when the scattering measurements or their upper limits become available. 
The uncertainty of $f_{\rm d}$ was not taken into account because the observational error of $f_{\rm d}$ is about 5$\%$ \citep{Cordes2022}, which is negligibly small compared with the error of DM$_{\rm cosmic}$ ($\sim$10\%-100\% in our sample).

We note that the $\tau$-DM$_{\rm host}$ relation is based on pulsar observations in the Milky Way, which is an edge-on galaxy. The different inclination angles to FRBs might affect the relation for FRBs.  This point can be improved by FRBs in nearby galaxies that may be studied in the future \citep[e.g.,][]{BURSTT}. 
In addition, the $\tau$-DM$_{\rm host}$ relation might not apply for FRB hosts because it is specific to Earth’s location in the Milky Way. Therefore, we note that there is no guarantee that this exact relation holds true for FRBs at different locations within different galaxy types of FRB hosts (see Sec \ref{Method:AFG} for a more detailed discussion and future work).

\subsection{Galaxy catalogs with photo-$z$}
\label{galaxy cat}
We used the photometric redshift (photo-$z$) galaxy catalogs of WISE-PS1-STRM \citep[WISE $\times$ PS1, hereafter;][]{Beck2022} and WISE $\times$ SuperCOSMOS \citep[WISE $\times$ SCOS, hereafter;][]{Bilicki2016} for the following reasons. 
WISE $\times$ PS1 encompasses photo-$z$ of flux-limited galaxy samples up to $z\sim$0.8 \citep{Beck2022}, which reasonably covers the redshift range of the majority of the FRB host samples (see Table \ref{FRBtable} and Section \ref{FRB} for the selection criteria).
Recently, the Deep Synoptic Array increased the FRB samples that were localized in the northern hemisphere with a host identification \citep{Law2023}. 
Therefore, WISE $\times$ PS1 makes a large fraction of FRB host samples available for a calculation of the number density.
WISE $\times$ SCOS incorporates flux-limited galaxy samples up to $z\sim$0.3 \citep{Bilicki2016}. 
Its whole-sky coverage allowed us to use FRBs in the northern and southern hemispheres \citep{Gordon2023, Bhandari2022}.

\subsection{Galaxy number density}
\label{density}
The projected foreground galaxy number densities ($\delta$) were measured around localized FRBs (see Table \ref{FRBtable} for FRB samples and Section \ref{FRB} for the selection criteria) as detailed below.
We used photo-$z$ galaxy catalogs of WISE $\times$ PS1 \citep{Beck2022} and WISE $\times$ SCOS (\citep{Bilicki2016} ; see Section \ref{galaxy cat} for details of the galaxy catalogs).
We applied a Vega magnitude cut for $W1 < 16.8$ mag to the WISE $\times$ PS1 galaxy catalog to construct flux-limited galaxy samples because the completeness of the catalog\footnote[1]{\url{https://wise2.ipac.caltech.edu/docs/release/allsky/expsup/sec6_3a.html}} significantly decreases at $W1 > 16.8$ mag. 
For each FRB host that spatially overlapped WISE $\times$ PS1, foreground galaxy samples were selected within a square region of 100$\times$100 Mpc$^{2}$ around the FRB. The selection criterion was photo-$z < $ $z_{\rm FRB}+\delta z_{\rm photo}$, where $z_{\rm FRB}$ is the spectroscopic redshift of the FRB host galaxy, and $\delta z_{\rm photo}$ is the uncertainty of the photo-$z$ for the galaxy samples.
Due to the error of the photo-$z$, the procedure sometimes invites high-$z$ galaxies into the foreground, while at other times, low-$z$ galaxies recede to the background. When the density is similar in the foreground and background of the FRB, this method returns the same number density, without a strong dependence on the size of the photo-$z$ error.
To take the variation of $\delta z_{\rm photo}$ at different redshifts and sky positions in the WISE $\times$ PS1 catalog into account, we adopted the median value of $\delta z_{\rm photo}$ of galaxies within $100\times100$ Mpc$^{2}$ scale {at photo-$z < z_{\rm FRB}+\sigma_{z}$ for each FRB, where $\sigma_{z}$ is the error of the photo-$z$. 
We adopted $\sigma_{z}=0.0287$ for the overall accuracy of the photo-$z$ in the entire WISE $\times$ PS1 catalog \citep{Beck2022}. 
As for WISE $\times$ SCOS, we used the overall photo-$z$ accuracy of $\sigma_{z} = 0.033(1+z_{\rm FRB})$ as $\delta z_{\rm photo}$ \citep{Bilicki2016}.

To calculate the projected galaxy number density, the number of galaxies was counted within a circular aperture centered at the position of each FRB host galaxy with a projected radius of 1 Mpc.
The aperture size of 1 Mpc was calculated by the angular diameter distance at the redshift of each FRB.  We chose this aperture radius in order to mimic the typical halo scale in a typical galaxy cluster.

Because fainter galaxies are not detected in a shallower survey or at a higher redshift, the absolute value of the galaxy number density depends on the galaxy survey depth and redshift.
To remove this effect, $100,000$ random apertures in the field of each FRB (100$\times$100 Mpc$^{2}$ scale centered on the FRB location) were investigated to compute reference densities with the same aperture radius.
We used the standard deviation of the reference densities to normalize the galaxy number density.
The density increment was measured relative to the peak of the histogram of the normalized reference densities for each FRB sample.
Therefore, the peak of the histogram, that is, $\delta=0$, represents the cosmic average of the galaxy number densities.
When $\delta>0$, the FRB lies in a region with a high foreground-galaxy density (hereafter, high-density FRB). 
Alternatively, when $\delta\leq0$, the FRB lies in a region with a low foreground-galaxy density (hereafter, low-density FRB).

The same procedure as described above was applied to FRB samples that spatially overlap WISE $\times$ SCOS.
Fig. ~\ref{fig1} shows examples of the histograms that depict the density increments of random apertures and FRB sight lines.

\begin{figure}
    \centering
    \includegraphics[width=\columnwidth]{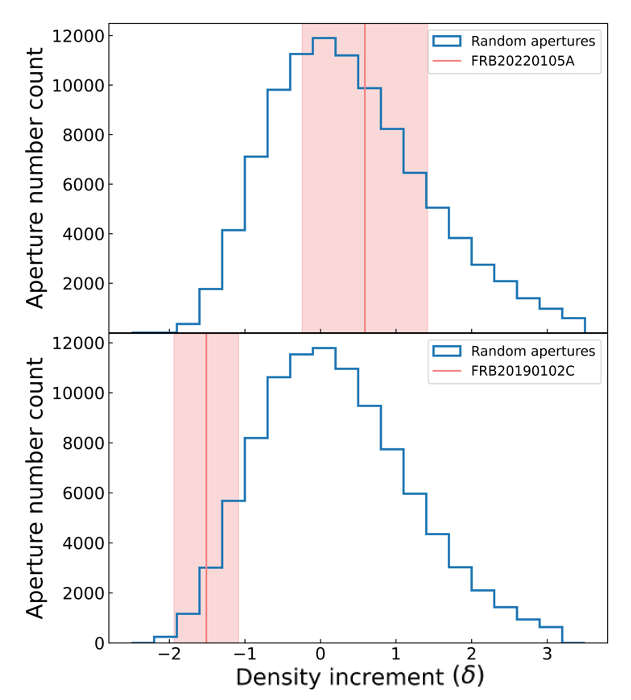}
    \caption{
Example histograms of density increments of random apertures and FRB sight lines.
Panels (a) and (b) show examples of FRBs in high and low galaxy-number-density environments, respectively, calculated by using WISE $\times$ SCOS samples.
In each panel, the blue histogram demonstrates the density increment of random apertures, and the vertical red line shows the density increment of the FRB sight line. The $\pm1\sigma$ errors are shown by the red shaded region.
    }
    \label{fig1}
\end{figure}

\subsection{FRB sample selection}
\label{FRB}
The FRB host samples were selected by the following selection criteria for WISE $\times$ PS1 (WISE $\times$ SCOS):
\begin{enumerate}
  \item[(1)] The spectroscopic redshift is available with a host identification
  \item[(2)] The FRB host galaxies are within the sky coverage of WISE $\times$ PS1 (WISE $\times$ SCOS)
  \item[(3)] $|b|>20$ deg
  \item[(4)] $z<0.8$ ($z<0.3$)
  \item[(5)] The star formation rate (SFR) of the host galaxy is in the range of $-1 \leq {\rm log}_{10}{\rm (SFR/M_{\odot}~yr^{-1})} \leq 1 $.
\end{enumerate}
(1) is necessary for the DM-$z$ relation.
(2) is necessary for the analysis of the galaxy number densities.
By criterion (3), we used only the FRB samples at a Galactic latitude $|b|>20$ deg because we found that the spatial distribution of galaxies at latitude $|b|<$ 20 deg is significantly inhomogeneous as a result of the strong dust extinction and/or the contamination of stars in the Milky Way disk.
We selected the FRB samples at $z<0.8$ ($z<0.3$) for WISE $\times$ PS1 (WISE $\times$ SCOS) to ensure the completeness of the galaxy catalog (4). 
The empirical relation we used to derive DM$_{\rm host}$ (Equation ~\ref{eq:tau}) is based on the plasma state in a Milky Way-like galaxy, but we assumed a systematically stronger scattering of FRBs \citep{CHIME2019scattering, Chawla2022} than in the Milky Way case by introducing a factor of 3--6 (see Section \ref{Method:DM} for details). 
The star formation rate is an indicator of the strength of the density fluctuation in the ISM. 
We applied an SFR cut (5) to the FRB samples to employ the empirical $\tau$-DM relation of Equation \ref{eq:tau}. 
Since the SFR of the Milky Way is $\textrm{SFR}_{\rm MW} = 2.0 \pm 0.7$ M$_{\odot}$ yr$^{-1}$ \citep{MWSFR}, we excluded samples with significantly deviating SFRs from SFR$_{\rm MW}$ beyond a range of one order of magnitude. 
The excluded samples are FRB 20190520B (SFR $= 0.04^{+0.04}_{-0.02} $ $\textrm{M}_{\odot}~{\rm yr}^{-1} $), FRB 20210117A (SFR $= 0.02^{+0.01}_{-0.01} $ $\textrm{M}_{\odot}~{\rm yr}^{-1}$), and FRB 20210410A (SFR $= 0.03^{+0.03}_{-0.01} $ $\textrm{M}_{\odot}~{\rm yr}^{-1} $) \citep{Gordon2023}.

In addition to the above criteria, FRB 20220509G was excluded from our samples because there might be scattering in the circumgalactic medium \citep{Conner2023}, which contradicts our assumption that the scattering preferentially occurs in host galaxies. 
The selected FRB samples for WISE $\times$ PS1 (WISE $\times$ SCOS) are summarized in Tables \ref{FRBtable} and \ref{DMhost table}.

\begin{table*}
    \centering
    \caption{Basic observational data of the FRB samples with SFR$_{100Myr}$ values and uncertainties.}
    \label{FRBtable}
    \renewcommand{\arraystretch}{1.1}
    \begin{tabular}{ccccccc}
        \hline
        \hline
        FRB & $z_{\rm FRB}$ & DM & SFR$^{a}$ & PS1$^{b}$ & SCOS$^{c}$ & Reference$^{b}$\\
 		    &  & ($\rm pc$ $\rm cm^{-3}$) & $M_{\odot}$/yr & &  &  \\
        \hline
        20190102C & 0.291    & 364.4   & $0.40^{+0.31}_{-0.11}$           & n  & y  & (1)(2)(3)(8) \\
        20190523A & 0.66     & 761     & $0.09^{+0.03}_{-0.03}$           & y  & n  & (3)(4)(5)\\
        20190608B & 0.1178   & 339.5   & $7.03^{+1.43}_{-1.15}$           & y  & y  & (1)(2)(5)(8) \\
        20190714A & 0.2365   & 504.1   & $1.89^{+1.22}_{-0.72}$           & y  & y  & (3)(6)(8)\\
        20191001A & 0.234    & 507.9   & $18.28^{+17.24}_{-8.95}$         & n  & y  & (3)(6)(7)(8)\\
        20200430A & 0.1608   & 380     & $0.11^{+0.06}_{-0.04}$           & y  & y  & (3)(6)(8)\\
        20200906A & 0.3688   & 577.8   & $4.93^{+3.46}_{-2.34}$           & y  & n  & (5)(8)\\
        20210320C & 0.2797   & 384.8   & $3.51^{+2.44}_{-1.45}$           & y  & y  & (8)(9)(8)\\
        20210603A & 0.1772   & 500.147 & $0.24^{+0.06}_{-0.06}$           & y  & y  & (10)\\
        20211127I & 0.0469   & 227     & $35.83^{+1.02}_{-1.46}$          & n  & y  & (8)(11)\\
        20211212A & 0.0707   & 209     & $0.73^{+0.62}_{-0.39}$           & y  & y  & (8) \\
        20220105A & 0.2785   & 580     & $0.42^{+0.31}_{-0.19}$           & y  & y  & (8) \\
        20220310F & 0.477958 & 462.24  & $4.25^{+2.45}_{-1.67}$           & y  & n  & (13)\\
        20220418A & 0.622    & 623.25  & $29.41^{+18.28}_{-10.70}$        & y  & n  & (13)\\
        20220914A & 0.1139   & 631.28  & $0.72^{+0.02}_{-0.15}$           & y  & y  & (13)(14)\\
        20220920A & 0.158239 & 314.99  & $4.10^{+1.09}_{-1.00}$           & y  & y  & (13)\\
        20221012A & 0.284669 & 441.08  & $0.15^{+0.04}_{-0.03}$           & y  & y  & (13) \\
        \hline
    \end{tabular}
    \begin{minipage}{\textwidth}
        \footnotesize
        $^{a}$ The star formation rate of the FRB host galaxy.\\
        $^{b}$ The columns (y$=$yes and n$=$no) show whether the FRB sample was used in the WISE $\times$ PS1 (PS1) or WISE $\times$ SCOS (SCOS).\\
        $^{c}$ References: (1) \cite{Macquart2020}, (2) \cite{Day2020}, (3) \cite{Cordes2022}, (4) \cite{Ravi2019}, (5) \cite{Bhandari2022}, (6) \cite{Heintz2020}, (7) \cite{Bhandari2020}, (8) \cite{Gordon2023}, (9) \cite{Sammons2023}, (10) \cite{Cassanelli2024}, (11) \cite{Glowacki2023}, (12) \cite{Bhandari2020}, (13) \cite{Law2023}, (14) \cite{Conner2023}.
    \end{minipage}
\end{table*}

\begin{table*}
	\centering
	\caption{Observed scattering, central observational frequency, and derived DM quantities along with the $DM_{\rm igm}$ values.}
	\label{DMhost table}
	\renewcommand{\arraystretch}{1.1}
	\begin{tabular}{cccccccccc}
		\hline
        \hline
		FRB & $\tau_{\rm obs}$ & $\tau_{\rm err}^{+}$ & $\tau_{\rm err}^{-}$ & $\nu^{a}$ & DM$_{\rm host}$$^{b}$ & DM$_{\rm host_{\rm err}}^{+}$ & DM$_{\rm host_{\rm err}}^{-}$ & DM$_{\rm cosmic}$$^{b}$ & reference$^{c}$ \\
		      & ($\rm ms$) & ($\rm ms$) & ($\rm ms$) & ($\rm GHz$) & ($\rm pc$ $\rm cm^{-3}$) & ($\rm pc$ $\rm cm^{-3}$) & ($\rm pc$ $\rm cm^{-3}$) & ($\rm pc$ $\rm cm^{-3}$) &  \\
            \hline
            20190102C & 3.3 & 0.041 & 0.041 & 1.27 & 147 & 48 & 70 & $125^{+63}_{-49}$ & (1)(2)(3) \\
            20190523A & 1.4 & 0.2 & 0.2 & 1 & 359 & 124 & 187 & $458^{+111}_{-80}$ & (3)(4) \\
            20190608B & 3.3 & 0.02 & 0.02 & 1.27 & 511 & 165 & 244 & $-194^{+217}_{-149}$ & (1)(2)(5) \\
            20190714A & $<2$ & N/A & N/A & 1.2725 & $<491$ & N/A & N/A & $215^{+135}_{-136}$ & (3)(6) \\
            20191001A & 3.3 & 0.02 & 0.02 & 0.824 & 257 & 86 & 129 & $171^{+121}_{-86}$ & (3)(6)(7) \\
            20200430A & $<15$ & N/A & N/A & 0.865 & $<371$ & N/A & N/A & $140^{+110}_{-108}$ & (3)(6) \\
            20200906A & N/A & N/A & N/A & N/A & 100 & 50 & 50 & $416^{+40}_{-40}$ & (5) \\
            20210320C & 0.247 & 0.006 & 0.006 & 0.842 & 151 & 50 & 73 & $178^{+57}_{-45}$ & (8)(9) \\
            20210603A & 0.155 & 0.003 & 0.003 & 0.6 & 69 & 24 & 35 & $352^{+32}_{-39}$ & (10) \\
            20211127I & N/A & N/A & N/A & N/A & 100 & 50 & 50 & $40^{+50}_{-50}$ & (11) \\
            20211212A & N/A & N/A & N/A & N/A & 100 & 50 & 50 & $24^{+50}_{-49}$ & (8) \\
            20220105A & N/A & N/A & N/A & N/A & 100 & 50 & 50 & $427^{+42}_{-42}$ & (8) \\
            20220310F & 0.02164 & 0.00022 & 0.00022 & 1.4 & 239 & 77 & 115 & $204^{+77}_{-57}$ & (13) \\
            20220418A & 0.0928 & 0.003 & 0.003 & 1.4 & 352 & 115 & 172 & $320^{+105}_{-75}$ & (13) \\
            20220914A & $<0.08$ & N/A & N/A & 1.4 & $<266$ & N/A & N/A & $405^{+81}_{-81}$ & (13)(14) \\
            20220920A & 0.078 & 0.015 & 0.015 & 1.4 & 271 & 96 & 144 & $-9^{+123}_{-87}$ & (13) \\
            20221012A & $<1.29$ & N/A & N/A & 1.4 & $<541$ & N/A & N/A & $123^{+144}_{-143}$ & (13) \\
		\hline
	\end{tabular}
        \begin{minipage}{\textwidth}
$^{a}$ $\nu$ is the observed frequency of FRB in GHz when the scattering time ($\tau_{\rm obs}$) and its error ($\tau_{\rm err}^{+}$ and $\tau_{\rm err}^{-}$) are measured. \\
$^{b}$ DM$_{\rm host}$ and DM$_{\rm cosmic}$ are derived quantities in this work (see Sec \ref{DM excess} for more information.)
$^{c}$ Reference for $\tau_{\rm obs}$ and $\nu$: $(1)$ \cite{Macquart2020}, $(2)$ \cite{Day2020}, $(3)$ \cite{Cordes2022}, $(4)$ \cite{Ravi2019}, $(5)$ \cite{Bhandari2022}, $(6)$ \cite{Heintz2020}, $(7)$ \cite{Bhandari2020}, $(8)$ \cite{Gordon2023}, $(9)$ \cite{Sammons2023}, $(10)$ \cite{Cassanelli2024}, $(11)$ \cite{Glowacki2023}, $(12)$ \cite{Bhandari2020}, $(13)$ \cite{Law2023}, $(14)$ \cite{Conner2023}.
    \end{minipage}
\end{table*}

\section{Results}
\subsection{Correlation between $\Delta$DM$_{\rm cosmic}$ and the density increment}
\label{DM-density}
Fig. ~\ref{fig2} shows the Macquart relation \citep{Macquart2020}, that is, DM$_{\rm cosmic}$ as a function of redshift with the data points from our FRB samples (Table ~\ref{FRBtable}). 
Fig. ~\ref{fig3}(a) and Fig.~\ref{fig3}(b) illustrate the relation of the density increment ($\delta$) and the deviation of  $\textrm{DM}_{\rm cosmic}$ ($\Delta $DM$_{\rm cosmic}$). 
We found a clear positive correlation between the density increment and $\Delta $DM$_{\rm cosmic}$.
The statistical significance and strength of the correlation were investigated by the weighted Pearson coefficient $p$-value test. 
To take the uncertainties of $\Delta$DM$_{\rm cosmic}$ and the density increment into account, we employed Monte Carlo simulations, which were conducted by incorporating random errors that follow Gaussian probability density functions. 

The observational uncertainties of $\Delta$DM$_{\rm cosmic}$ and the density increment were adopted as the standard deviations of the Gaussian functions.
The uncertainty of $\Delta$DM$_{\rm cosmic}$ is asymmetric toward the upper and lower bounds. 
We therefore adopted for each data point two Gaussian probability density functions (PDFs) representing the upper and lower bounds of $\Delta$DM$_{\rm cosmic}$. 
The right (left) side of the $\Delta$DM$_{\rm cosmic}$ PDF can be expressed by the right (left) half of a Gaussian with a standard deviation of the $\Delta$DM$_{\rm cosmic}$ upper (lower) bound.
The Monte Carlo simulations were performed 100,000 times, and we derived 100,000 $p$ values of the Pearson coefficient by taking the errors into account. 

$\Delta$DM$_{\rm cosmic}$ was modeled as a function of $\sqrt{z_{\rm FRB}}$ \citep{Macquart2020}. It indicates the higher significance of the deviation at the lower redshift for a given $\Delta$DM$_{\rm cosmic}$. 
To take this effect into account, we weighted the data points with $1/\sqrt{z_{\rm FRB}}$ while conducting the Pearson coefficient $p$-value test.
The results of the statistical test are shown in Fig. \ref{fig4}.
In Fig. \ref{fig4}, the median $p$ values of the correlation between the density increment and $\Delta$DM$_{\rm cosmic}$ are $p=0.012$ and 0.032, and the median Pearson coefficients are 0.6 and 0.6 for the WISE $\times$ PS1 and WISE $\times$ SCOS samples, respectively. 
We also tested the statistical significance and strength of the correlation without weighting. 
The $p$ values are $p=0.027$ and 0.038, and the Pearson coefficients are 0.6 and 0.6 for WISE $\times$ PS1 and WISE $\times$ SCOS, respectively. This confirms that weighting does not affect our result significantly.

The results with the WISE $\times$ PS1 and WISE $\times$ SCOS samples both indicate statistically significant correlations 
between $\Delta$DM$_{\rm cosmic}$ and the foreground galaxy number density around the FRBs within the uncertainties.

\begin{figure}
    \centering
    \includegraphics[width=\columnwidth]{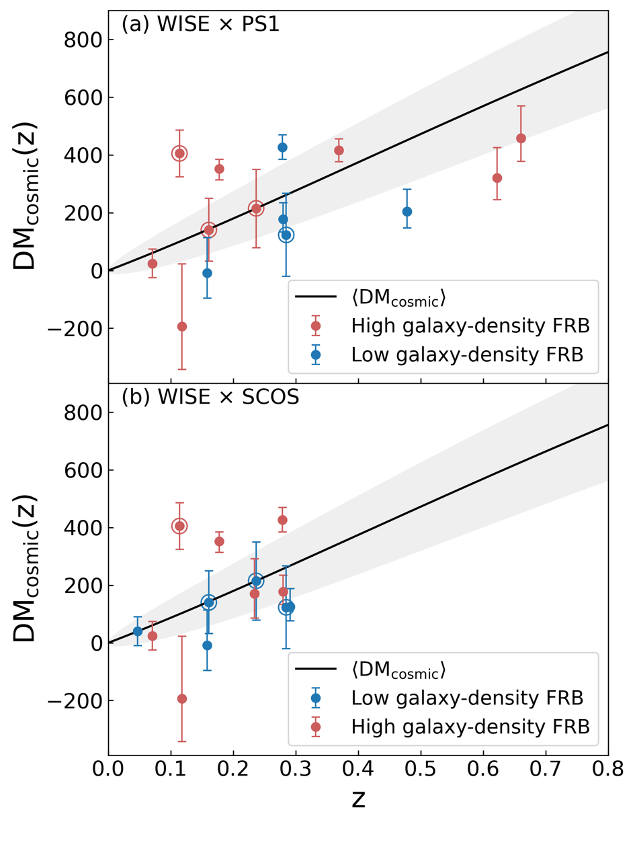}
    \caption{
Panels (a) and (b): Macquart relation , i.e., DM$_{\rm cosmic}$ (pc cm$^{-3}$) as a function of redshift, with the WISE $\times$ PS1 and SCOS $\times$ WISE FRB samples, respectively. 
The high-density ($\delta>0$) FRB samples are denoted as red data points, and the low-density ($\delta\leq0$) FRB samples are represented by blue data points (see Section \ref{density} for the details of the galaxy number-density calculation).
The FRB samples with upper limits on $\tau$ are marked with open circles on the data points. 
The solid black line shows the theoretical prediction of the average DM$_{\rm cosmic}$, $\langle$DM$_{\rm cosmic}\rangle$,  as a function of $z$ by adopting the \texttt{Planck18} cosmology \citep{Planck18} and the fraction of the cosmic baryons in the diffuse gas, $f_{\rm d}=$0.85. 
The gray shaded region shows the expected scatter of DM$_{\rm cosmic}$ due to the cosmic variance \citep{McQuinn2014, Cordes2022}.
    }
    \label{fig2}
\end{figure}

\begin{figure}
    \centering
    \includegraphics[width=\columnwidth]{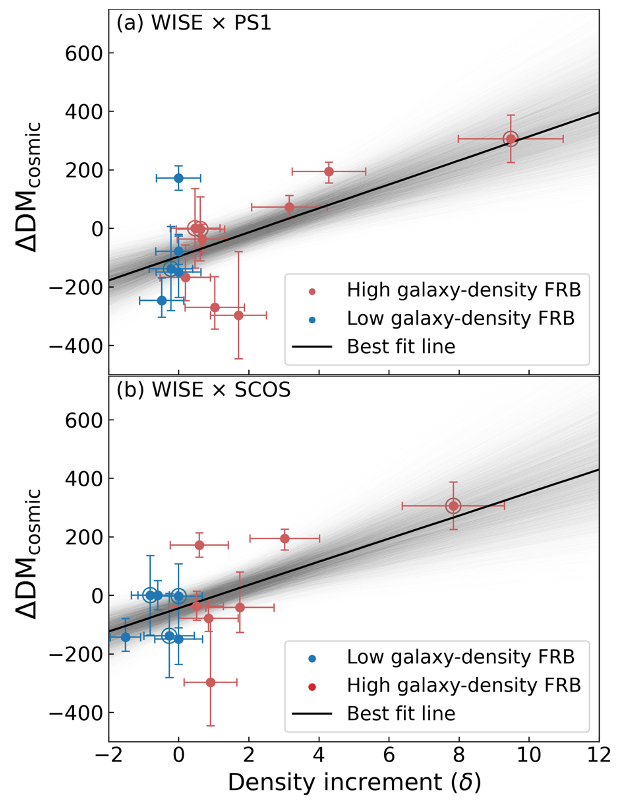}
    \caption{
Panels (a) and (b): Deviation of DM$_{\rm cosmic}$ ($\Delta $DM$_{\rm cosmic}$) (pc cm$^{-3}$) from the theoretical curve (solid black line in Fig. ~\ref{fig2}) as a function of the density increment with WISE $\times$ PS1 and SCOS $\times$ WISE FRB samples, respectively. 
The coloring of the data points is the same as Fig. \ref{fig2}. 
The thin gray lines indicate the individual best-fit linear functions to the data points, randomized with the uncertainties following the Gaussian probability density functions by the Monte Carlo simulations. 
The solid black line indicates the median of the best-fit linear functions.
    }
    \label{fig3}
\end{figure}

\begin{figure*}
    \centering
    \includegraphics[width=2\columnwidth]{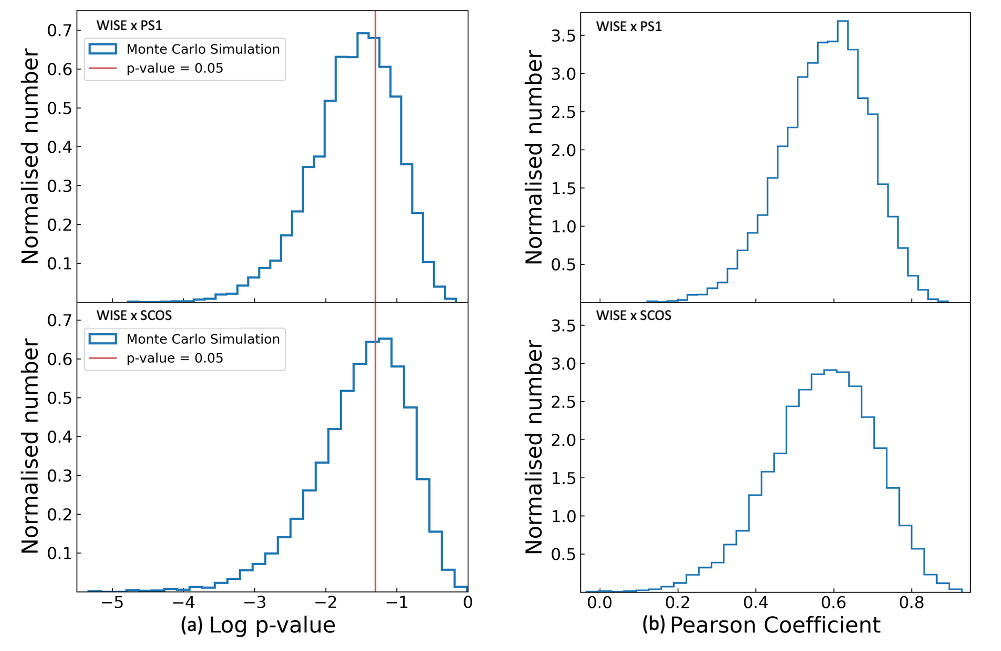}
    \caption{
Left panels: Histograms of the $p$ values derived from the Monte Carlo simulations to test the statistical significance of the correlation between the deviation of DM$_{\rm cosmic}$ ($\Delta $DM$_{\rm cosmic}$) and the density increment (Fig. ~\ref{fig3}).
The top and bottom panels show the WISE $\times$ PS1 and WISE $\times$ SCOS samples, respectively. 
The conventionally adopted significance threshold, $p$ value $=0.05$, is shown as the vertical red line. 
Right panels: Same as the left panels, but for the Pearson coefficients derived from the Monte Carlo simulations.  
    }
    \label{fig4}
\end{figure*}

\begin{figure}
    \centering
    \includegraphics[width=\columnwidth]{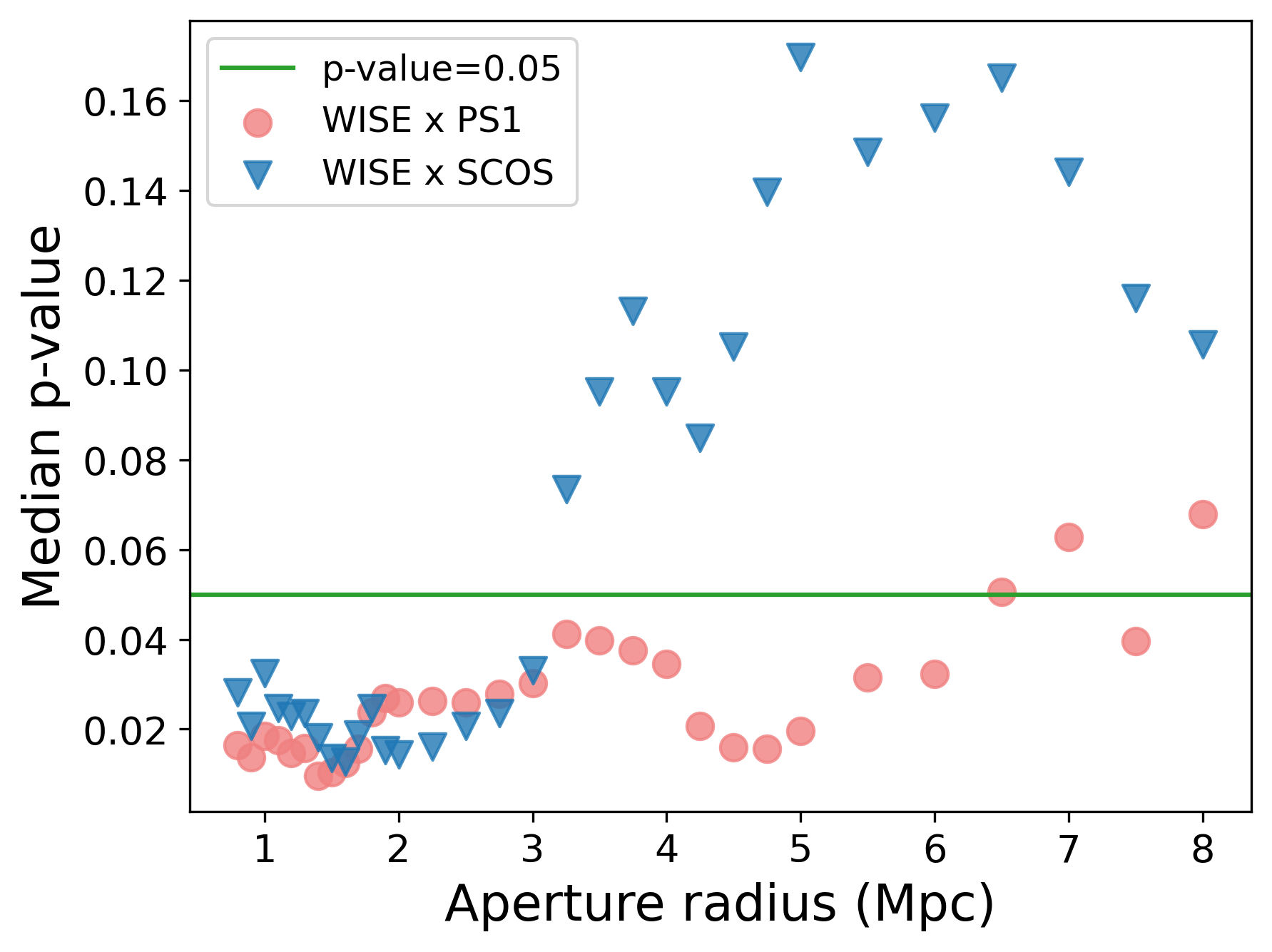}
    \caption{
Median $p$-value of the correlation between the deviation of DM$_{\rm cosmic}$ ($\Delta $DM$_{\rm cosmic}$) and the density increment as a function of the aperture radius for calculating the galaxy number density. 
The red data points indicate the results with the WISE $\times$ PS1 catalog, and the blue data points show the $p$-value results for the WISE $\times$ SCOS. 
The horizontal green line indicates the threshold of the statistical significance of the correlation, i.e., $p= 0.05$.
    }
    \label{fig5}
\end{figure}

\subsection{Typical baryonic fluctuation scale}
\label{Methods:radius} 
We employed a radius of the aperture of 1 Mpc to calculate the foreground galaxy number density because the typical scale of galaxy-cluster halos is 1 Mpc. 
We also investigated the different assumptions on the aperture radius.
The median $p$ value of the correlation between $\Delta$DM$_{\rm cosmic}$ and the density increment was calculated with the same procedure as described in Section \ref{DM-density}, but with different aperture radii. 
Fig. \ref{fig5} shows the median $p$ value as a function of the aperture radius.
The correlation becomes insignificant ($p$ values $> 0.05$) when a radius of $>$ 6 Mpc and $>$ 3 Mpc is adopted for WISE $\times$ PS1 and WISE $\times$ SCOS, respectively. 
These characteristic spatial scales were evaluated at the FRB redshifts.  
The corresponding physical scale is smaller for foreground galaxies at lower $z$ than the FRB redshift.
Therefore, the characteristic scale is interpreted as an upper limit on the typical baryonic fluctuation scale.
Considering the conservative upper limit placed by the WISE $\times$ PS1 sample, our result indicates a characteristic baryonic fluctuation scale of $\lesssim$ 6 Mpc at $z < 0.8$. 
It is unclear why WISE$\times$SCOS drops to a low-significance correlation at 3-6 Mpc. Possible reasons include that the larger photo-$z$ errors of WISE$\times$SCOS might lose sensitivity at larger apertures where the density contrast is weaker. Spectroscopic data would be ideal for accurately measuring the correlation scale.

\section{Discussion}
\subsection{Comparison with previous case studies}
Several case studies argued for possible associations between the DM$_{\rm cosmic}$ excess from the $\langle$DM$_{\rm cosmic}$$\rangle$ and the foreground galaxy mapping \citep[e.g.,][]{Conner2023,Lee2023,Wu2023} as follows. 
The FRBs that are embedded in massive galaxy clusters, FRB 20220914A and FRB 20220509G, exhibit a DM$_{\rm cosmic}$ excess that is likely contributed by the intracluster medium (ICM) \citep{Conner2023}. 
Our samples include FRB 20220914A, which shows a high-density foreground galaxy environment with $\delta$ = 9.5 $\pm$ 1.5  and $\delta$ = 7.4 $\pm$ 1.4  for the WISE $\times$ PS1 and WISE $\times$ SCOS, respectively, and a DM$_{\rm cosmic}$ excess of 305$_{\rm -81}^{\rm +81}$ pc cm$^{-3}$. This confirms the robustness of our method.
The possible DM$_{\rm cosmic}$ excess of FRB 20190520B was reported in line with the multiple foreground galaxy clusters and galaxy groups along the line of sight to the FRB \citep{Lee2023}. 
About 20-30 FRBs were detected with the Canadian Hydrogen Intensity Mapping Experiment (CHIME) that intersect foreground galaxy halos for $<$ 80 Mpc. 
A stack analysis of these FRBs showed a marginal 1$\sigma$ to 2$\sigma$ DM excess \citep{Wu2023}.

Moreover, the statistical angular cross-correlation between the first CHIME/FRB catalog and galaxy catalogs was studied \citep{Rafiei-Ravandi2021}, and a potential association between high galaxy-number density regions and FRBs at $0.3 \lesssim z \lesssim 0.5$ was reported. 
This analysis was different from this work because we considered both low and high galaxy-number density regions and used localized FRB, but the previous analysis used nonlocalized FRBs.

In addition to observational studies, \citet{Walker2023} used IllustrisTNG 300-1 simulations to show that filaments of large-scale structures dominate DM$_{\rm cosmic}$ in void regions. The voids and filaments correspond to low- and high-density environments, respectively. Therefore, the result of \citet{Walker2023} is qualitatively consistent with the positive correlation between $\Delta $DM$_{\rm cosmic}$ and the density increment we found here. The slope of the correlation may depend on the hydrodynamic simulations that are adopted. It would also depend on the feedback models of supernovae and AGNs \citep[e.g.,][]{Medlock+2024}, which are yet to be known, because the strength of the feedback controls the amount of plasma outside galaxies. Therefore, the quantitative comparison between the observed slope and the slope that is predicted in cosmological simulations is beyond the scope of this paper.

Previous observational studies \citep{Conner2023,Lee2023,Wu2023,Rafiei-Ravandi2021} investigated particular galaxy clusters or nearby galaxy halos and focused on the association between the DM$_{\rm cosmic}$ excess and a high-density environment. 
In contrast to these works, we quantitatively took all of the density environments of foreground galaxies into account, including the deficit/excess of DM$_{\rm cosmic}$ and low/high-density regions in a general manner. 
Our analysis is therefore more general in describing the cosmological variation in the medium outside the galaxies and presents statistical evidence of the cosmological fluctuation of missing baryons. 

Our findings indicate that outside galaxies, baryons are preferentially found in regions with a higher galaxy density, and that they preferentially avoid regions with a lower galaxy density on scales of $\lesssim$ 6 Mpc.
This has been predicted by the theoretical structure formation scenarios \citep{Walker2023}, but was hard to prove observationally because a direct probe of the missing baryons is lacking.
The unique capability of DMs of cosmological FRBs uncovered not only the missing baryons \citep{Macquart2020}, but also their cosmological fluctuation in this work.

\subsection{Where are the missing baryons?}
\label{Methods:caution2}
A key question in the missing baryon problem is how they are distributed 
outside the galaxies because they may come from very distinct environments, that is, from the hot intergalactic medium, from clusters, or from halos. However, it is difficult to distinguish them in the current analysis because the photo-z error is large and the impact parameter is accordingly uncertain. The DM excess/deficit in our analysis therefore includes all of the possible components, including the intergalactic medium, the intracluster medium, and the circumgalactic medium. Separating these components is essentially important and might be done by future spectroscopic surveys of the FRB sight lines \citep[e.g.,][]{flimflam}.

\subsection{$\Delta$DM$_{\rm cosmic}$ as a function of the density increment}
\label{Methods:caution}
We note that the data point with the highest galaxy number density in Fig. \ref{fig3} dominates the statistical significance of the correlation. When this data point is removed, the best-fit linear function still shows a positive correlation, although the statistical significance becomes lower with $p$ values higher than 0.05 for most cases of the Monte Carlo realizations. However, we may be able to improve this with more localized FRB samples in the future.

\subsection{DM$_{\rm cosmic}$ derivation}
\label{Method:DM}
In Fig. \ref{fig2}, some of the derived DM$_{\rm cosmic}$ values are negative. 
The negative values of DM$_{\rm cosmic}$ are mathematically expected to occur within the statistical errors of  DM$_{\rm cosmic}$.
When we force DM$_{\rm cosmic}$ to be positive in our analysis, a systematic bias might be raised in calculating DM excess/deficit. 
To avoid these systematics, we allowed this negative DM$_{\rm cosmic}$ in our analysis within the errors. 

We also tested $f_{\rm scale}=6$, which corresponds to the scattering in an intervening clump in a host galaxy \citep{Cordes2016}.
In this case, the scattering in FRB host galaxies is even stronger than the Milky Way for a given DM$_{\rm host}$.
The same analysis as for $f_{\rm scale}=3$ was performed with Monte Carlo simulations.
The median $p$ values of the correlation between the deviation of DM$_{\rm cosmic}$  and density increment are $p=0.006$ and 0.009 for WISE $\times$ PS1 and WISE $\times$ SCOS, respectively. 
We conclude that the different assumptions on the scattering-DM$_{\rm host}$ relation do not affect our results significantly.

We assumed that the scattering and foreground galaxy number densities are independent in adopting Equation \ref{eq:tau}.
However, if the observed scattering and foreground galaxy number densities were anticorrelated, DM$_{\rm host}$ would be overestimated (underestimated) for low (high) galaxy-number densities.
This effect might mimic the correlation between $\Delta$DM$_{\rm cosmic}$ and the density increment. To test this point, we investigated the hypothetical anticorrelation between $\tau_{\rm 1GHz}$ and the density increment for our samples with the Pearson coefficient $p$-value test. The error of $\tau_{\rm 1GHz}$ and the density increment were taken into account by Monte Carlo simulation. 
We found no significant anticorrelation between the scattering and galaxy number density, with a median $p$ value of 0.5 and a Pearson coefficient of -0.2 for the samples in both catalogs.

\subsection{Alternative approach to estimating DM$_{\rm host}$}
\label{Method:AFG}

We used the empirical relation between $\tau_{\rm 1GHz}$ and DM$_{\rm host}$ established based on Galactic pulsars to estimate DM$_{\rm host}$ (Equation \ref{eq:tau}).
An alternative approach is important to address this point.
One possible approach is the \lq Couldlet model\rq\ \citep{Cordes2022}, which is the physically motivated model of a scattering screen in an FRB host galaxy.
This model describes the relation between scattering ($\tau_{\rm 1GHz}$), which is assumed to occur in a host galaxy, and the square of DM$_{\rm host}$ with a coefficient of $A_{\tau}\tilde{F} G$.
$A_{\tau}\tilde{F} G$ is the product of three parameters, where $A_{\tau}$ is the parameter that shows that empirical estimates for the scattering time are related more closely to the $e^{-1}$ time than to the mean, $\tilde{F}$ describes the degree of the density fluctuation of plasma in a host galaxy, and $G$ is the geometric factor that approaches unity when the distance between an FRB and the observer is much larger than the thickness of the scattering screen, that is, the case for extragalactic FRBs.
This model parameterization can be applied to the observed spectroscopic redshifts of FRB samples to optimize the DM-derived redshifts \citep{Cordes2022} and hence DM$_{\rm host}$. 
However, in this fitting process, the model allows an extremely wide dynamic range of the $A_{\tau}\tilde{F} G$ parameter as a prior assumption, which is a variation of three orders of magnitude. 
This means that DM$_{\rm host}$ in this approach has the freedom to change by about $\sqrt{3} \sim 1.7$ orders of magnitude for a given observed scattering.
This extreme freedom of DM$_{\rm host}$ is much larger than any DM excess/deficit discussed in this work, which is typically a factor difference in DM. 
Therefore, any true variation in the DM excess/deficit can easily be marginalized into DM$_{\rm host}$, which makes this approach less sensitive to detecting a DM excess/deficit.
To fully use this approach by narrowing down the prior $A_{\tau}\tilde{F} G$ range, we need a better understanding of the detailed physical process of scattering \citep{Cordes2022}, which is yet to be known and is beyond the scope of this paper. 
Therefore, we used the empirical relation instead of the physically motivated modeling.
In addition, pulsars in a face-on configuration will be useful in understanding the empirical relation between DM and scattering for face-on low-mass FRB host galaxies. However, within the Milky Way, more pulsars are seen toward the Galactic disk, that is, in an edge-on configuration, than face-on. Therefore, pulsars in the two Magellanic clouds would be useful to address this point because they are seen in a much less edge-on configuration than pulsars in the Milky Way.
We leave this subject to future work.

\section{Conclusion}
\label{Conclusion}
A unique method for probing missing baryons in the Universe are FRBs. 
The Macquart relation \citep{Macquart2020} indicates that a significant number of missing baryons reside outside galaxies.
The next key question is the statistical fluctuation of this relation, which is anticipated to be caused by the variation in cosmic structures.
Previous studies reported possible associations between the excess in DM$_{\rm cosmic}$ over its cosmic average and regions with a high galaxy-number density that intersect the lines of sight to FRBs \citep[e.g.,][]{Rafiei-Ravandi2021,Conner2023, Lee2023, Wu2023}.
However, these studies only focused on galaxies with a high number density.
To fully understand the cosmological fluctuation of the missing baryons, it is important to take the excess/deficit in DM$_{\rm cosmic}$ and the high/low galaxy-number densities into account in a systematic manner.

To address this point, we employed (nearly) all-sky galaxy catalogs, including WISE $\times$ PS1 \citep{Beck2022} and WISE $\times$ SCOS \citep{Bilicki2016}, to investigate the foreground galaxy-number densities around 14 and 13 localized FRBs, respectively.
We found a positive correlation between the deviation of DM$_{\rm cosmic}$ from its cosmic average and foreground galaxy-number density for the WISE $\times$ PS1 and WISE $\times$ SCOS catalogs.
The correlation is strong and statistically significant, with median Pearson coefficients of 0.6 and 0.6 and median $p$ values of 0.012 and 0.032 for WISE $\times$ PS1 and WISE $\times$ SCOS, respectively.
The statistical significance of the correlation is enhanced with density aperture sizes of $\lesssim$ 6 Mpc and $\lesssim$ 3 Mpc for WISE $\times$ PS1 and WISE $\times$ SCOS, respectively.

Our findings unveil the statistical evidence of cosmological fluctuation when localized FRBs are used at a characteristic fluctuation scale of $\lesssim$ 6 Mpc.
The unique capability of DMs of cosmological FRBs uncovered not only the missing baryons \citep{Macquart2020}, but also their cosmological fluctuation in this work.

\begin{acknowledgements}
We would like to express our gratitude toward the anonymous referee for the comprehensive and thoughtful review of our manuscript. The author thanks NTHU $\times$ NCHU cosmology group, Dr. Alvina Y. L. On and Prof. J. Xavier. Prochasca for their insightful comments and suggestions. We acknowledge the Taiwan Astronomical ObserVatory Alliance (TAOvA) of the summer student internship in partial financial support of this research. We thank the Wide Field Astronomy Unit (WFAU) at the Institute for Astronomy, Edinburgh, for the WISE $\times$ SCOS photo-$z$ catalog. T-YH thanks the National Science and Technology Council of Taiwan Undergraduate Research Grant for University Students 112-2813-C-005 -025 -M. TG acknowledges the support of the National Science and Technology Council of Taiwan through grants  112-2112-M-007 -013, and 112-2123-M-001 -004 -. 
TH thanks the support of the National Science and Technology Council of Taiwan (NSTC) through grants 110-2112-M-005 -013 -MY3, 113-2112-M-005-009-MY3, 110-2112-M-007-034-, and 113-2123-M-001-008- and the support of the Ministry of Education Republic of China (Taiwan) through a grant 113RD109. 
SY acknowledges the support from NSTC through grant numbers 113-2811-M-005-006- and 113-2112-M-005-007-MY3. SH acknowledges the support of The Australian Research Council Centre of Excellence for Gravitational Wave Discovery (OzGrav) and the Australian Research Council Centre of Excellence for All Sky Astrophysics in 3 Dimensions (ASTRO 3D), through project number CE17010000 and CE170100013, respectively. 
\end{acknowledgements}

% \bibliography{ref.bib}  

\begin{appendix} 
\section{Malmquist's bias of galaxy selection}

\label{malmquist}
In order to check whether Malmquist’s bias has a significant impact on this work, we have checked survey-volume difference between higher-z cone ($0 <z_{\rm galaxy} < z_{\rm FRB} + \delta z_{\rm photo}$) and lower-z cone ($0 <z_{\rm galaxy} < z_{\rm FRB} - \delta z_{\rm photo}$) of the FRB due to the photo-z error from the galaxy catalog. At the median redshift of z = 0.2785 for PS1 samples and z = 0.1775 for SCOS samples, the volume ratio between the higher-z cone and the lower-z cone is 1.3 and 1.7 for PS1 and SCOS samples, respectively. This selection thus increases the number of galaxies by this factor. However, when evaluating galaxy density, we apply the same selection criteria to 100000 random apertures and the FRB aperture to ensure a fair comparison.

\section{Further test of the correlation between $\Delta$DM$_{\rm cosmic}$ and density increment}
\label{Appendix}

We have also conducted the weighted Pearson Coefficient $p$-value test for the correlation between $\Delta$DM$_{\rm cosmic}$ and density increment (1) using the FRB FRB samples only with |b|>40 and (2) including FRB 20220509G in our original sample for both WISE $\times$ PS1 and WISE $\times$ SCOS. The same methods are used and described in Section ~\ref{DM-density}. With the |b|>40 criteria, FRB 20220914A (|b|$\sim$ 26) is removed from the sample. This FRB resides in the highest density increment. As discussed in Section 4.3, this FRB dominates the statistical significance of the correlation. Therefore, this test is almost identical to the one in Section 4.3. The median p-value of the Monte Carlo Simulation is greater than 0.05 for both WISExPS1 and WISExSCOS samples. Apart from the above test,  we conducted another test by including FRB20220509G in our analysis. In this case, both of the WISE $\times$ PS1 and WISE $\times$ SCOS samples show no significant correlation (median $p$-value > 0.05 in Monte Carlo simulation). The scattering happening in the circumgalactic medium stated in \citet{Conner2023} leads to the DM deficit even though it is in the massive galaxy cluster.

\end{appendix}

\end{document}